# A new sulfur bioconversion process development for energy- and space-efficient secondary wastewater treatment

Chu-Kuan Jiang[a], Yang-Fan Deng[a,b], Hongxiao Guo[a], Guang-Hao Chen[a,b,*], Di Wu[a,c,d,*]

a. Department of Civil and Environmental Engineering, Water Technology Centre, Hong Kong Branch of Chinese National Engineering Research Centre for Control & Treatment of Heavy Metal Pollution, The Hong Kong University of Science and Technology, Hong Kong, China.

b. Wastewater Treatment Laboratory, Fok Ying Tung Graduate School, The Hong Kong University of Science and Technology, Guangzhou, China.

c. Centre for Environment and Energy Research, Ghent University Global Campus, Incheon, Republic of Korea.

d. Department of Green Chemistry and Technology, Ghent University, and Centre for Advanced Process Technology for Urban Resource Recovery, Ghent, Belgium.

*Corresponding authors:

Di Wu, Centre for Environment and Energy Research, Ghent University Global Campus (di.wu@ghent.ac.kr); Guang-Hao Chen, Department of Civil and Environmental Engineering, The Hong Kong University of Science and Technology (ceghchen@ust.hk).

**Highlights**

- The ERATO was developed and validated in a moving-bed biofilm reactor.
- The Thio-S/TdS-S ratio reached as high as 38–73% with influent DO above 2.7 mg/L.
- Significant thiosulfate promotion by oxygen supply was confirmed.
- $SO_4^{2-}$ and microbial activities are also critical for thiosulfate formation.
- BSR metabolism was regulated by oxygen, contributing to thiosulfate production.

**Abstract**

Harvesting organic matter from wastewater is widely applied to maximize energy recovery via biogas production from anaerobic digestion or power generation from incineration; however, it limits the applicability of secondary treatment for acceptable effluent discharge into surface water bodies. To turn this bottleneck issue into an opportunity, this study developed oxygen-induced thiosulfatE production duRing sulfATe reductiOn (EARTO) to provide efficient electron donor for wastewater treatment. Typical pretreated wastewater was synthesized with chemical oxygen demand of 110 mg/L, sulfate of 50 mg S/L, and varying dissolved oxygen (DO) and was fed into a moving-bed biofilm reactor (MBBR). The MBBR was operated continuously with a short hydraulic retention time of 40 min for 349 days. The formation rate of thiosulfate reached 0.12–0.18 g S/(m$^2$·d) with a high thiosulfate-S$_{produced}$/TdS-S$_{produced}$ ratio of 38–73% when influent DO was 2.7–3.6 mg/L. The sludge yield was 0.23–0.29 gVSS/gCOD, much lower than it in conventional activated sludge processes. Then, batch tests and metabolism analysis were conducted to confirm the oxygen effect on thiosulfate formation, to characterize the roles of sulfate and microbial activities, and to explore the mechanism of oxygen-induced thiosulfate formation in ERATO. Results examined that oxygen supply promoted the thiosulfate-S$_{produced}$/TdS-S$_{produced}$ ratio from 4% to 24–26%, demonstrated that sulfate and microbial activities were critical for thiosulfate production, and indicated that oxygen induces thiosulfate formation through two pathways: 1) direct sulfide oxidation ($S^{2-}$ + $O_2$ → $S_2O_3^{2-}$), and 2) indirect sulfide oxidation, sulfide is first oidized to $S^0$ (dominant) which then reacts with sulfite derived from oxygen-regulated biological sulfate reduction ($S^0$ + $SO_3^{2-}$ → $S_2O_3^{2-}$). The proposed compact ERATO process, featuring high thiosulfate production and low sludge production, supporting space- and energy-efficient secondary wastewater treatment.

# 1. Introduction

Municipal wastewater treatment plants (WWTPs) consume a staggering amount of energy, accounting for 3–4% of global electricity consumption (IEA, 2022). Lowering the energy consumption and resultant greenhouse gas emission of WWTPs is crucial. A practical approach is to energy-efficiently maximize the recovery of the organic matter (energy) from wastewater. Anaerobic treatment (Petropoulos et al., 2017), high-rate activated sludge (Jimenez et al., 2015), and chemically enhanced primary treatment (CEPT) processes (Diamantis et al., 2013; Harleman and Murcott, 1999) were intensively investigated and applied. Take CEPT as an example, it captures 65–75% of organics in sewage (Maktabifard et al., 2018; Ødegaard, 2016), and then potentially 2.5–2.9 KWh/m$^3$ of energy could be recovered (McCarty et al., 2011). However, the low organics concentration in the effluent limits the application of conventional nitrification/denitrification for further nitrogen removal (Wan et al., 2016).

Nowadays, sulfate concentration of 20–180 mg S/L is commonly found in wastewater because of the seawater intrusion (Widlansky et al., 2020), the application of seawater toilet flushing (Wu et al., 2016), and the dosage of sulfate-based coagulants (e.g., $Al_2(SO_4)_3$, $Fe_2(SO_4)_3$, and polyferric sulfate) in CEPT for treatment of upstream drinking water or wastewater (Pikaar et al., 2014; Renault et al., 2009; Wang et al., 2009). Considering the sulfate source and the chemical oxygen demand (COD) of 60–150 mg/L coexisted in the pretreated effluent (Kfouri and Kweon, 2003; Maktabifard et al., 2018), the energy-efficient sulfur bioconversion associated processes are feasible alternatives for treating organics-insufficient mainstream wastewater (Cui et al., 2019a). For the successful development of these processes, e.g., sulfur-driven denitrification or partial denitrification and anammox (Cui et al. 2019a, Deng et al. 2022), sulfide and thiosulfate are important electron carriers and can be produced in biological sulfate reduction (BSR).

Compared to sulfide, thiosulfate drives autotrophic denitrification with a 4.6 times higher rate (Cardoso et al., 2006) thus leading to an high denitrification activity of 6.72 kg N/(m$^3$·d) with an low hydraulic retention time (HRT) of 15 min (Qian et al., 2021). Moreover, 0.1 mM of sulfide would inhibit denitrification and the 50% inhibitory concentration was 0.01 mM for anammox (Cojean et al., 2020; Russ et al., 2014). Thiosulfate has no inhibition effect on denitrification and anammox until the concentration reaches 46.0 mM and 4.9 mM, respectively (Liu et al., 2022; Lubina et al., 1996). Therefore, thiosulfate formation in BSR is beneficial for accelerating of the downstream nitrogen removal and facilitating the development of biological sulfur-based processes. However, the produced thiosulfate was generally recognized as a byproduct of BSR with a low thiosulfate-$S_{produced}$/TdS-$S_{produced}$ (Thio-S/TdS-S) ratio of 0–17% (Hao et al., 2013).

Different from seldomly found in BSR without oxygen presence, thiosulfate is commonly found together with sulfide and oxygen. For instance, in marine or freshwater sediments where BSR occurs, a thiosulfate peak was found together with sulfide in the oxic/anoxic interface (Jørgensen, 1990). Specifically, 68–78% of sulfide was oxidized to thiosulfate in the anoxic environments. Meanwhile, in sulfide-rich gas treatment, applying continuous oxygen supply in the sulfide-oxidizing bioreactors, 17–35% of oxidized sulfide was stably converted into thiosulfate in the bioreactors (Klok et al., 2012; Van den Bosch et al., 2008). The findings of thiosulfate accumulation in co-existence of sulfide and oxygen enlightened that thiosulfate promotion in BSR by introducing oxygen is possible. However, most related studies are focused on the inhibition of the BSR to enhance anaerobic digestion, recovery of the element sulfur during BSR (as microaeration applied in treating sulfate-rich wastes), and the versatility of sulfate-reducing bacteria (SRB) under the presence of oxygen (Cypionka, 2000; Nguyen and Khanal, 2018; Xu et al., 2012). It remains unclear whether oxygen supply could contribute to thiosulfate accumulation in BSR.

This study aims to develop an oxygen-induced thiosulfatE production duRing sulfATe reductiOn (EARTO), supporting space- and energy-efficiently secondary wastewater treatment. A moving-bed biofilm reactor (MBBR), fed with synthetic pretreated wastewater and varied amount of oxygen supply, was operated for 349 days to 1) investigate the performance of this system; 2) identify the functional microbial community in the ERATO; 3) evaluate thiosulfate formation in ERATO; and 4) explore the mechanism and metabolism of thiosulfate conversion.

## 2. Materials and methods

### 2.1 Reactor setup and operation

A MBBR (Fig. S1 in SI 1) with 1 L of working volume was fabricated using acrylic plastic. This reactor was homogeneously mixed at 120 rpm using a mechanical stirrer. Sulfate reducers were cultivated using carriers (BioChip 30$^{TM}$, Germany) with depth, diameter, and high specific surface area of 1.1 mm, 30 mm, and 5500 m$^2$/m$^3$, respectively. The carriers were taken from our previous sulfate-reducing reactor studying the treatment of saline wastewater with a sulfate concentration of 180 mg S/L (Leung, 2019). The filling ratio was set as 40%, corresponding to approximately 300 carriers. The MBBR was fed with synthetic wastewater, mimicking CEPT effluent, at a constant flow rate of 1.5 L/h, which led to a short nominal HRT (40 min). The synthetic wastewater contained 110 mg COD/L, 30 mg $NH_4^+$-N/L, and 50 mg $SO_4^{2-}$-S/L. The bioreactor temperature was maintained at 27–28 °C (Table 1) using a water bath. The pH of the bioreactor was controlled at 7.0–7.3. The DO of the bioreactor was <0.5 mg/L. No sludge was purposely wasted from the system. During the operation, the influent and effluent samples were collected 2–4 times a week to measure nitrogen and sulfur compounds.

To investigate the effect of oxygen on thiosulfate production, the operation of the continuous reactor was divided into Phase I~III based on different DO concentrations in the influent. In Phase I (day 1–day 136), the feasibility of the treatment of synthetic wastewater via BSR was examined with an influent DO of 2.7 ± 0.5 mg/L. Phase II lasted from day 137 to day 270 and was characterized by a low influent DO concentration of 0.5 ± 0.3 mg/L (with nitrogen gas sparging). In Phase III (day 271–day 349), influent DO concentration increased again to 3.6 ± 0.8 mg/L.

**2.2 Batch tests**

A total of two groups of batch tests (BTs), A and B, were conducted to reveal sulfur conversion in the bioreactor, as summarized in Table 2. In BT-A, five batch tests were designed to characterize the anaerobic sulfate reduction (BT-A1), examine the effects of oxygen on thiosulfate production (BT-A2), explore the roles of sulfate concentration and microbial activities in thiosulfate formation (BT-A3~A5). The other group BT-B containing three batch tests (BT-B1~B3) was conducted further investigate the mechanism of oxygen-induced thiosulfate formation in ERATO.

BT-A1 was performed to determine the activities of BSR based on a medium containing COD of 110 mg/L and sulfate concentration of 50 mg S/L (the control experiment). BT-A2 (experimental test) involved air addition (oxygen supply) was conducted to investigate the effect of oxygen on thiosulfate formation. During BT-A2, 22.5 mL of air was added to the batch bottles under anaerobic conditions; specifically, 4.5 mL of air was added at 30, 60, 90, 120, and 150 min. To verify whether thiosulfate formation was based on BSR, different initial sulfate concentrations of 30 mg S/L and 80 mg S/L were designed in BT-A3 and A4 respectively with other conditions set the same with BT-A2 at first. Then sterilized carriers were applied in BT-A5 with other initial conditions kept the same as in BT-A2. The carriers were sterilized through autoclaving (twice) at 120 °C for 30 min (Van de Graaf et al., 1995). In BT-B, BT-B1 and B2 were designed to analyze whether oxygen contributed to thiosulfate formation via direct chemical sulfide oxidation (Klok et al., 2012). BT-B1 and B2 were performed with initially 25 mg S/L of total dissolved sulfide (TdS) and 50 mg/L of COD mediated by sterilized or non-sterilized carriers, and 4.5 mL of air was added at 0, 30, 60, 90, and 120 min (22.5 mL in total). At last, BT-B3 was conducted to investigate the role of produced $S^0$ in thiosulfate formation. BT-B3 was conducted under the same initial conditions as BT-A1, except that in the former, the medium was spiked with 2.5 g $S^0$-S/L of sublimed sulfur.

For each BT, a bottle with a 205 mL working volume was applied. The headspace and liquid phase were sparged with nitrogen gas for 30 min to remove oxygen. Then, 27 carriers were taken from the MBBR (resulting in a specific surface area of 1.0 $m^2$/L for each batch bottle) and washed three times with deoxygenated ultrapure water. Nitrogen gas flushing was applied again for 20 min after the carriers were placed into the batch bottle to remove residual oxygen. The stock solutions (as described in SI 2) were applied to achieve the required concentrations of sulfate, COD, and TdS. The temperature was controlled at 30 °C using a gas bath shaker at a shaking rate of 180 rpm. The initial pH of each BT was controlled at 7.5 ± 0.2 through the introduction of 1 N HCl or 1 N NaOH solution. Bulk liquid samples with 5–10 mL volume each were taken every 30–90 min from the batch bottles. A gas bag (1 L) was filled with nitrogen gas and connected to the batch bottle to maintain a constant pressure and anaerobic environment during the batch experiments and sampling period. All the tests were performed in duplicate.

## 2.3 Chemical analysis

Before chemical analysis, liquid samples were first dosed with 0.1 mL of 5 N NaOH solution and then filtered with a 0.22 μm Millipore$^{TM}$ syringe. TdS was measured via the methylene blue method (APHA, 2005). Afterward, 0.1 mL of 1 mol/L ZnCl$_2$ solution was added to the samples for TdS stabilization (ZnS precipitation), and the samples were filtered again using 0.22 μm Millipore$^{TM}$ syringe filters (Keller-Lehmann et al., 2006). Then, the sulfate (SO$_4^{2-}$-S) and thiosulfate (S$_2$O$_3^{2-}$-S) concentrations were measured using an ion chromatographer HIC-20A super (Shimadzu, Japan) equipped with a conductivity detector and an IC-SA2 analytical column. Total dissolved organic carbon was analyzed with a total organic carbon (TOC) analyzer TOC-5000A (Shimadzu, Japan). COD, total suspended solids (TSS), and volatile suspended solids (VSS) was measured according to APHA.(2005). The conversion factor between TOC and COD was determined as 2.66 via measurements. The attached total solids (ATS, g/L) and attached volatile solids (AVS, g/L) of the MBBR biofilm were measured according to the procedure reported by (Cui et al., 2018). The biofilm sample and the yellow complex attached to the wall inside the MBBR were collected on day 349 and analyzed via Raman microscopy (Renishaw RM 3000 Micro-Raman system). The measurement conditions were set the same as in (Cui et al., 2019b). The DO concentration, pH, and temperature of the reactor were measured using the YSI Pro Plus multiparameter meter (YSI, USA). Dissolved methane (CH$_4$) was measured according to (Guisasola et al., 2008) .

## 2.4 Calculation and statistical analysis

The kinetic reaction rates in BSR were calculated as described in SI 3 using the data obtained from the MBBR performance and the results of BTs. The mass balance for sulfur was calculated as follows: ΔS$_2$O$_3^{2-}$-S + ΔTdS-S)/|ΔSO$_4^{2-}$-S| × 100%. The molar ratio of O$_{2\text{-supply}}$/SO$_4^{2-}{}_{\text{initial}}$ was used in batch tests, O$_{2\text{-supply}}$ and SO$_4^{2-}{}_{\text{initial}}$ refer to the accumulated supply of oxygen and potential sulfide production, respectively. The applied ratio was modified from the molar ratio of $O_2$/TdS was applied for regulating sulfide-oxidizing process (Van den Bosch et al., 2008).

The significant differences in thiosulfate formation of the MBBR between the different phases or the BTs were analyzed via the least significant difference test (LSD, $p < 0.05$). The correlation between thiosulfate formation and influent DO was characterized by calculating the Pearson's Product Moment Correlation Coefficient in Origin 9.0. The range of coefficient value is -1 to 0 or 0 to +1, and 0 refers to no correlation; -1 and +1 suggest a perfect negative and positive correlation, respectively (Prion and Haerling, 2014).

**2.5 Microbial community analysis**

The biofilm samples were taken from the MBBR on days 136, 270, and 349 for microbial community analysis. Three carriers were taken from the bioreactor to obtain every biofilm sample. The universal forward primer 515F (5′-GTGCCAGCMGCCGCGG-3′) and the reverse primer 907R (5′-CCGTCAATTCMTTTRAGTTT-3′) were used for bacteria amplification, targeting the V4 and V5 regions of the bacterial 16S gene (Quince et al., 2011). Redundancy analysis (RDA) was conducted to investigate the relationship between influent DO concentration change and microbial community shifts. Moreover, the obtained 16S rRNA sequences were analyzed by the Phylogenetic Investigation of Communities by Reconstruction of Unobserved States 194 (PICRUSt) method (Langille et al., 2013) for further investigation of the thiosulfate conversion pathway. As this study only focused on the predicted enzymes, the prediction type of Kyoto Encyclopedia of Genes and Genomes Orthologs is chosen. The raw data were uploaded to the National Center for Biotechnology Information with the BioProject accession number PRJNA903667.

## 3. Results and discussion

### 3.1 Performance of the bioreactor

#### 3.1.1 Development of BSR

The lab-scale MBBR was operated for 349 days with the influent DO concentration range from 0.5–3.6 mg/L. The long-term performance of the bioreactor is shown in Fig. 1 and Table 3. In Phase I (days 0–136), with an average influent DO of 2.7 ± 0.5 mg/L, the surface-specific BSR rate reached 0.34 g S/(m$^2$·d) on day 100. At the end of Phase II (days 247–270), the BSR rate was increased to 0.38 g S/(m$^2$·d), and it was maintained at 0.42 g S/(m$^2$·d) during Phase III (days 271–349), indicating that the BSR activity in the MBBR was stable for over 100 days and that the influent DO (0.5–3.6 mg/L) did not impair the BSR activity. Accordingly, the TdS generation rate was reached and maintained at 0.32 g S/(m$^2$·d) from Phase II to Phase III. During 349 days of cultivation, biofilm was enriched with the ATS of 0.96–1.39 g/L and AVS of 0.82–1.20 g/L, respectively. Meanwhile, the VSS in effluent was 13.5–17.3 mg/L. Including the accumulated and the observed biomass yield, total sludge yield was 0.23–0.29 gVSS/gCOD. The yield is much lower than 0.45 g VSS/g COD of activated sludge production (Ekama and Wentzel, 2020).

With relatively low sulfate of 48.1–54.0 mg S/L in influent, the volumetric BSR rate in this MBBR reached and maintained at 0.92 kg S/(m$^3$·d). The BSR rate in this study was higher than that of 0.30–0.73 kg S/(m$^3$·d) in granular sulfate-reducing reactors supplied with glucose/acetate and high sulfate of >150 mg SO$_4^{2-}$-S/L (Hao, 2014). During the operation, the sulfur mass balance was 105%–119%. Meanwhile, the volumetric organics removal rate was stabilized at 1.91 kg COD/(m$^3$·d) in Phase III. The ratio between removed organics and reduced sulfate was 2.1–2.2 (mg COD/mg SO$_4^{2-}$-S), slightly greater than the reported value of 2.0 (Lens et al., 1998). It suggests that the influent DO may consume the electrons in organics. As competition and co-existence of sulfate-reducing organisms and methanogens always exist, the $CH_4$ production was also detected in the effluent with a low concentration of 0.02 mg/L on average, accounting for <0.1% COD consumption. This finding was in accordance with that methanogenesis was negligible at a COD/$SO_4^{2-}$-S ratio of <5.1 (Choi and Rim, 1991).

**3.1.2 The relationship between thiosulfate production and influent DO**

The different concentrations of influent DO were applied during Phase I~III to explore its effects on thiosulfate formation. In Phase I with an average influent DO of 2.7 mg/L as Fig. 2a shown, thiosulfate concentration in the effluent reached 10.3 ± 3.7 mg S/L. Notably, the Thio-S/TdS-S ratio reached 73% which is incredibly higher than the typical ratio of 0–17%. In Phase II, when the mean value of influent DO concentration declined to 0.5 mg/L, the thiosulfate concentration was decreased to 4.9 ± 3.1 mg S/L with a Thio-S/TdS-S ratio of 24%, close to the ratio of 0–17% in literature (Hao et al., 2013). Then, with the influent DO concentration increased to 3.6 mg/L in Phase III, the thiosulfate was increased and maintained at a concentration of 7.5 ± 2.8 mg S/L with the Thio-S/TdS-S ratio stabilized at 38%. During Phase I~III, the thiosulfate production rate was in range of 0.08–0.18 g S/($m^2$·d). A high thiosulfate formation rate of 0.12–0.18 g S/($m^2$·d) was found in Phase I and III with high influent DO (2.7–3.6 mg/L). These results indicate that oxygen in the influent contributed to thiosulfate production in the MBBR. Then, Pearson's correlation was adopted to analyze the relationship between influent DO and thiosulfate formation, as shown in Fig. 2b. The obtained Pearson correlation coefficient of 0.57 disclosed a moderate positive correlation between thiosulfate formation and the influent DO. Furthermore, considering that oxygen could contribute to $S^0$ formation where BSR occurs (Xu et al., 2012), $S^0$ in the biofilm sample and yellow substance attached inside the wall of the MBBR were analyzed and confirmed via Raman (Fig. S2 in SI 4).

**3.2 Microbial community analysis**

Biofilm samples were taken from the MBBR on days 136, 270, and 349 to identify the functional bacteria at the phylum and genus levels. The relationships between the dynamics of the microbial community and the concentrations of influent DO and thiosulfate in different phases were analyzed.

As shown in Fig. 3a and b, three dominant phyla in the sludge sample were *Desulfobacterota* (formerly *Deltaproteobacteria*), *Proteobacteria,* and *Bacteroidota*, with relative abundances in the range of 18.7−38.8%, 12.9−33.6% and 14.4−23.2% respectively. Specifically, the abundance of *Desulfobacterota* (to which most of the recognized SRB species belong (Castro et al., 2000)) and *Proteobacteria* were 38.8% and 21.0%, separately on day 349. Meanwhile, the dominant SRB genera shifted from *Desulfobulbus* (2.1%) and *Desulfuromusa* (1.3%) in inoculated biofilm (Leung, 2019) to *Desulfobacter* (33.4%), *Desulfobulbus* (1.3%), and *Desulfomicrobium* (0.2%) in biofilm sample obtained in the MBBR after 136 days of operation. With low influent DO (0.5 ± 0.3 mg/L) in Phase II, the abundance of the most dominant SRB genera *Desulfobacter* decreased to 9.6% on day 270, then increased to 26.8% on day 349 when influent DO was back to 3.6 ± 0.8 mg/L. The abundances of other SRB bacteria, *Desulfobulbus* and *Desulfomicrobium*, continuously increased from 1.3% and 0.2% on day 136 to 4.9% and 1.0% on 349, respectively, irrespective of the varying DO in the influent. The abundance of *Desulfobacter* was considerably higher than those of other SRB genera during Phase I~III, indicating that *Desulfobacter* was the dominant sulfate-reducing bacteria for ERATO. Not surprisingly, sulfur-oxidizing bacteria (SOB) such as *Sulfurovum*, *Thioclava,* and *Thiothrix* (Ghosh and Dam, 2009; Nielsen et al., 2000) were found in our biofilm system. The total abundance of SOB was 1.4−10.2% during days 136−349. The existence of SOB is consistent with the observation of $S^0$ in the MBBR, and with the fact that SOB could coexist with SRB when provided oxygen (Xu et al., 2012).

Furthermore, the relationship between the abundance of enriched bacteria at genus level and the concentrations of influent DO and produced thiosulfate in different phases was further analyzed by RDA. Results (Fig. 3c) revealed a positive relationship between the abundance of *Desulfobacter* and the concentrations of the influent DO and produced thiosulfate. As *Desulfobacter* was able to utilize oxygen (Cypionka, 2000), the obtained positive relationship indicated that *Desulfobacter* was favoured by oxygen supply. Meanwhile, *Thioclava* and *Sulfurovum* were also positively related to thiosulfate concentrations during Phase I~III. These results suggested that *Desulfobacter, Thioclava,* and *Sulfurovum* were responsible for thiosulfate formation in ERATO.

### 3.3 Investigation of thiosulfate production in ERATO by batch test

#### 3.3.1 Anaerobic sulfate reduction: BSR activity and thiosulfate formation

BT-A1 with no oxygen supply was designed to investigate the rates of anaerobic sulfate reduction and thiosulfate formation. As shown in Fig. 4a, BSR reaction mainly occurred in the first 90 min with the initial sulfate and COD concentrations of 54.6 ± 3.2 mg S/L and 118.6 ± 1.7 mg/L, respectively. The surface-specific rates of BSR, TdS production, and COD consumption were 23.21 mg S/(m²·h), 20.47 mg S/(m²·h), and 50.80 mg/(m²·h), respectively, as calculated via linear regression. The linear coefficients of determination ($R^2$) for these three variations were higher than 98%. The specific sulfate reduction rate of 56.1 mg S/(gVSS·h) was considerably higher than the reported rate in the granular sludge system (~30 mg S/(gVSS·h) (Hao, 2014). This result demonstrated that the biofilm with high sulfate-reducing activity was developed in MBBR. After which, the concentrations of sulfate and organics were slowly decreased to 12.1 ± 2.5 mg S/L and 40.2 ± 10.6 mg/L separately at 180 mins. With the deoxygenated environment, no thiosulfate is produced in the first 150 mins. Only 1.5 ± 0.4 mg S/L of thiosulfate was found produced at 180 min with a Thio-S/TdS-S ratio of 4%. This is consistent with the phenomenon in the MBBR that low influent DO leads to low thiosulfate formation in Phase II. Meanwhile, no COD consumption in this period indicated that the thiosulfate was derived from sulfur conversion.

### 3.3.2 Oxygen supply promotes thiosulfate formation

The BT-A2 results demonstrated the effect of oxygen on BSR (Fig. 4b). The initial COD and sulfate concentrations were 112.0 ± 2.8 mg/L and 48.3 ± 0.5 mg S/L, respectively. Similar to the BT-A1, BSR is mainly carried out in the first 90 min. The air was supplied at 30 min and 60 min. Notably, 7.7 ± 0.5 mg S/L of thiosulfate was formed at 90 min with a Thio-S/TdS-S ratio of 26%, which is significantly higher than it found in BT-A1 ($p < 0.05$). This result is in accordance with the positive relationship between influent DO and thiosulfate formation in the MBBR. Therefore, the oxygen contribution to thiosulfate formation in BSR was further examined since oxygen was the only variant ($N_2$ in air belongs to inert gas). Within 90 mins, the rates of BSR, TdS production, and COD degradation were 22.08 mg S/($m^2$·h), 20.09 mg S/($m^2$·h), and 64.86 mg/($m^2$·h), respectively, comparable to the results obtained in BT-A1. During 90–180 mins with another three dosages of air, thiosulfate concentration was decreased first to 5.0 mg S/L at 120 min and then kept increased to 7.2 ± 0.7 mg S/L at the end. After which, the sulfate concentration maintained at around 15.2 mg S/L during 90–150 mins, indicating that BSR was stopped at 90 mins. Moreover, the sulfate increase (3.2 mg S/L) and TdS decrease (10.7 mg S/L) after 120 mins suggested the occurrence of sulfide oxidation (Van den Bosch et al., 2008), in consistence with the existence of SOB found in biofilm.

Molar ratio of $O_{2\text{-supply}}/SO_4^{2-}{}_{initial}$ was used to analyze the suitable amount of oxygen supply for thiosulfate formation. First of all, thiosulfate was produced and maintained between 5.0 and 7.7 mg S/L during 90–180 min with $O_{2\text{-supply}}/SO_4^{2-}{}_{initial}$ molar ratio being 0.26–0.66. Although the concentration is fluctuated with air dosage, the Thio-S/TdS-S ratio of 17–40% was much higher than it in BT-A1. On the other hand, the trends of reaction rates and compounds variation before 150 min in BT-A2 and BT-A1 were similar, indicating that oxygen did not adversely affect BSR under an $O_{2\text{-supply}}/SO_4^{2-}{}_{initial}$ molar ratio of 0.52. With $O_{2\text{-supply}}/SO_4^{2-}{}_{initial}$ increased to 0.63 after 150 min, the decrease in TdS (8.4 mg S/L) and the increase in sulfate (3.3 mg S/L) suggested the occurrence of TdS oxidation and the excess of oxygen supply. The results suggested that oxygen supply via microaeration with $O_{2\text{-supply}}/SO_4^{2-}{}_{initial}$ of 0.26–0.52 is suitable for promoting thiosulfate formation. Meanwhile, a similar suitable ratio of $O_{2\text{-supply}}/SO_4^{2-}{}_{initial}$ (0.22–0.44) was found in the extra batch test with a total air dosage of 35 mL, as presented in SI 5.

**3.3.3 Exploration of the roles of sulfate concentration and microbial activities**

To further confirm the contribution of sulfate and microbial activities on thiosulfate formation, three batch tests in total were performed (BT-A3~5). The effect of sulfate concentration was investigated by BT A3 and A4. As Fig. 4c shows, no thiosulfate production was observed in BT-A3 with initial sulfate of 26.8 ± 1.0 mg S/L during the whole batch. Maximum TdS production of 22.2 ± 1.1 mg S/L was obtained at 60 mins and then maintained at between 20.1–21.9 mg S/L during 90–180 mins, indicating that BSR ceased at 60 mins with residual sulfate of 7.3 ± 0.4 mg S/L. Compared with BT-A2, the only difference is low sulfate existence during 60–180 mins, indicating that sufficient sulfate is critical for thiosulfate formation. Sulfate concentration of below 9.6 mg S/L would limit both BSR rate and sulfate uptake rate (Habicht et al., 2005). Therefore, no formation of thiosulfate during 60–180 mins might be due to the lack of BSR activities (including activities of sulfate reduction and sulfate uptake). In line with the results, Thio-S/TdS-S ratio emerged and reached 24% at 150 mins in BT-A4 with sulfate beyond 43.1 mg S/L during the whole test. The results of BT-A2, A3, and A4 indicate that oxygen contributed to thiosulfate promotion with sulfate kept above 15.2 mg S/L.

BT-A5 with sterilized carriers was performed to exclude the contribution of unknown chemical reactions. As shown in (Fig. 4d), the concentrations of sulfate, TdS, thiosulfate, and COD were constant during the whole batch, indicating the successful sterilization of carriers. Under this circumstance, the non-occurrence of thiosulfate production confirmed that the contribution of chemical reaction to thiosulfate formation in ERATO was excluded (Van de Graaf et al., 1995). Therefore, microbial activities are responsible for ERATO. The sulfur mass balances for all BTs above were in range of 85%–105%.

### 3.4 Mechanisms of thiosulfate production in ERATO

To better understand how oxygen affects thiosulfate formation, BT-B1~B3 and PICRUSt were performed to explore possible mechanisms responsible for thiosulfate formation and conversion in ERATO.

### 3.4.1 Direct and indirect sulfide oxidation

In the bioreactor, the average oxidation-reduction potential during the operation was $-311 \pm 41$ mV. Biological and chemical sulfide oxidation can both occur with ORP in the range of $-265$ to $-420$ mV (Khanal and Huang, 2006; Klok et al., 2012). As the main products of sulfide oxidation were $S^0$, sulfate, and thiosulfate, the contribution of direct oxidation ($S^{2-}$ + $O_2$ → $S_2O_3^{2-}$) and indirect oxidation with $S^0$ formation to thiosulfate formation were investigated in BT-B1, B2, and B3.

Both chemical and biological oxidation of sulfide were investigated in BT-B1 and B2 with sterilized and alive carriers, respectively. The results are shown in Fig. 5a and 5b. In BT-B1 with an initial TdS concentration of 23.2 ± 0.4 mg S/L, the abiotic sulfide oxidation rate was 10.39 mg S/(m$^2$·h) during 0−120 min. After which, the residual TdS (2.5 ± 1.3 mg/L) was all consumed at 150 min. And during 60−150 min, 4.1–5.4 mg S/L of thiosulfate was produced, demonstrating that thiosulfate could be formed via abiotic sulfide oxidation. Meanwhile, with alive carriers and the initial TdS concentration of 25.3 ± 2.1 mg S/L in BT-B2, the TdS oxidation rate was 10.91 mg S/(m$^2$·h) during 0−120 min. Maximumly 1.8 ± 0.6 mg S/L of thiosulfate, significantly lower than it in BT-B1 ($p < 0.05$), was formed at 120 min. In line with that, products of biological sulfide oxidation were mainly $S^0$ and $SO_4^{2-}$ (Ghosh and Dam, 2009; Van den Bosch et al., 2008). At 150 min, a thiosulfate decrease of 0.8 mg S/L was further observed, indicating bioconversion of thiosulfate. Compared to BT-A2, only 1.0–1.8 mg S/L of thiosulfate production in BT-B2 suggests that direct sulfide oxidation was not mainly responsible for thiosulfate formation in ERATO. Meanwhile, it is worth noting that in both BT-B1 and B2, the calculated $S^0$ concentrations were 20.7 mg S/L and 19.9 mg S/L, respectively, at the end of the tests. It indicates that $S^0$ was the main product of sulfide oxidation as reported in biological sulfide oxidation process (Van den Bosch et al., 2008). Therefore, the effect of $S^0$ on thiosulfate formation in BSR was further investigated in BT-B3 (Fig. 5c). With 2.5 g/L of $S^0$ added, thiosulfate concentration was found to reach 2.4 ± 2.0 mg S/L at 150 min and then increased to 2.7 ± 0.2 mg S/L at 180 min with a sulfur balance of 85%. Thiosulfate formation was significantly higher than it in BT-A1 (1.5 mg S/L) or in BT-B2 (1.0–1.8 mg S/L) ($p < 0.05$). This result demonstrates that $S^0$ contributed to thiosulfate formation and indicates that interaction between BSR and formed $S^0$ from indirect sulfide oxidation was mainly responsible for thiosulfate formation in ERATO.

To further understand thiosulfate conversion in EARTO, the average and maximum Thio-S/TdS-S ratios obtained in all BTs were summarized in Fig. 5d. From the results of BT-A1, A2, and A4, the average Thio-S/TdS-S ratio was found increased tremendously from 4% without $O_2$ supply to 24–27% with $O_2$ supply when initial sulfate concentration was higher than 48.3 mg S/L. Moreover, the 0% ratio obtained in both BT-A3 (low sulfate concentration) and BT-A5 (sterilized carriers) revealed the critical role of BSR in thiosulfate formation. The average ratios in BT-B1 and B2 were 18 ± 4% and 6 ± 1% respectively, the significantly lower ratio obtained in BT-B2 ($p < 0.05$) indicated the biological sulfide oxidation and biodegradation of thiosulfate. At last, the average ratio in BT-B3 was 1.5 times that in BT-B2, indicating that $S^0$ played a more important role in contributing to thiosulfate formation. The results of the maximum Thio-S/TdS-S ratios in all BTs also showed a similar trend with the change of average ratios.

### 3.4.2 Reaction between $S^0$ and sulfite

One way that $S^0$ can lead to thiosulfate formation is by reacting with sulfite in BSR. For instance, the absence or lack of DsrC, which could accelerate biological sulfite reduction, would slow down the sulfite reduction rate with sulfite accumulation, which then could contribute to thiosulfate formation (Santos et al., 2015). As the abundance of enzymes like DsrC would affect the thiosulfate formation, the oxygen effect on the expression of sulfur-related enzymes in biofilm were analyzed via PICRUSt. According to KEGG, enzymes related to sulfur conversion were summarized in Fig. 6. CysAPUW, SULP are sulfate transporters. Sat, AprAB, and DsrAB and DsrC are responsible for mediating sulfate reduction. PhsAC, TST, and GlpE transporters are capable of utilizing thiosulfate. Additionally, YchN is involved in sulfur oxidation, and SorA and SUOX can mediate sulfite oxidation. These enzymes were identified in the predicted results of the biofilm samples taken on day 136, 270 and 349.

Notably, with influent DO of 2.7–3.6 mg/L during Phase I and III, the predicted expression of DsrC, which determines the sulfite reduction rate, was found with 2 units. More units of DsrC (3) were obtained on day 270 with less oxygen supplied in Phase II (influent DO of 0.5 mg/L). Less DsrC units found with more oxygen supply suggested that oxygen inhibited DsrC expression. DsrC decrease greatly slows down sulfite reduction (sulfite accumulation) with the production of intermediates (e.g., $S^0$), thus forming thiosulfate by the reaction between sulfite and $S^0$ (Leavitt et al., 2015; Santos et al., 2015). Meanwhile, the same units of SorA (1) and SUOX (1) for sulfite oxidation found in three different phases indicated comparable sulfite consumption. Therefore, more sulfite would accumulate with oxygen supply, causing more thiosulfate formation ($SO_3^{2-} + S^0 \rightarrow S_2O_3^{2-}$). This is consistent with the conclusions obtained in sections 3.3.3 and 3.4.1 that both sulfate concentration (for sulfite provision) and $S^0$ affected thiosulfate formation.

### 3.4.3 Proposed thiosulfate conversion

For a better understanding of thiosulfate production in ERATO, the thiosulfate conversion routes were summarized in Fig. 7. In the perspective of thiosulfate production, oxygen functioned through sulfide oxidation and, more importantly, the reaction between formed $S^0$ and sulfite. As mentioned in sections 3.4.1 and 3.4.2, oxygen could lead to $S^0$ formation and sulfite accumulation through sulfide oxidation and BSR with upregulating DsrC, respectively. For biological sulfide oxidation, the YchN (uncharacterized protein in oxidation of intracellular sulfur) was found as the main sulfur oxidation enzyme with 0–16 units, instead of common sulfur-oxidizing enzymes like Sox complex, SQR, and FCSD in predicted results (Ghosh and Dam, 2009). From the perspective of thiosulfate degradation, oxygen could lead to thiosulfate oxidation mediated by GlpE with sulfite production. Meanwhile, PhsAC and TST responsible for reducing thiosulfate with sulfite formation were found. The total units of PhsAC and TST were 159, much higher than the units of GlpE (1), indicating that thiosulfate reduction is mainly responsible for thiosulfate conversion. Furthermore, sulfite production from both thiosulfate oxidation and reduction could support the recovery of thiosulfate through the reaction between sulfite and $S^0$.

## 3.5 Implications

With existence of oxygen, the conversion of sulfate, sulfide and thiosulfate in sediments was demonstrated by applying radiotracer experiments (Jørgensen, 1990). Based on which, this study reveals the role of oxygen in thiosulfate formation, facilitating a better understanding of sulfur conversion. Meanwhile, ERATO is demonstrated and developed, benefiting space- and energy-efficient secondary wastewater treatment. With 40 mins of HRT for this compact process, 38–73% of thiosulfate promotion could accelerate downstream autotrophic denitrification and reduce the demand of space by up to 50%. It should be noted that the rates of autotrophic and heterotrophic denitrification are comparable (Cui et al., 2019a), thiosulfate promotion is beneficial for the fastest nitrogen removal. Meanwhile, thiosulfate promotion with 28–42% of sulfide decrease would effectively alleviate the inhibition effect of sulfide on anammox. Therefore, it's beneficial to the development of sulfur-driven partial denitrification anammox for energy-efficient nitrogen removal (Deng et al., 2022). Moreover, the sludge production in this process was only 51–64% of it in conventional activated sludge process. Considering that sludge treatment takes up to 20–65% of energy consumption in WWTPs, energy consumption would be reduced by 6–29% in this process (Sid et al., 2017).

Nevertheless, this study is only focused on confirming the feasibility of this process. The long-term operation was limited to the COD and sulfate concentrations of 48.1–54.0 mg S/L and 109.8–118.4 mg/L, respectively, different from the fluctuated concentrations in real sewage (e.g., sulfate was in range of 20–180 mg S/L). For better guidance of the application, the effect of critical parameters like oxygen supply, COD/SO$_4^{2-}$, pH and T on thiosulfate promotion need to be investigated.

**4. Conclusion**

A new bioprocess ERATO significantly improved thiosulfate production was developed in this study, supporting space- and energy-efficiently secondary wastewater treatment of wastewater after organics capture. With 349 days of operation, the thiosulfate production rate reached 0.12−0.18 g S/(m$^2$·d) with a high Thio-S/TdS-S ratio of 38–73% when influent DO was 2.7–3.6 mg/L in the bioreactor. Meanwhile, sulfate-reducing bacteria (*Desulfobacter*) and sulfate-oxidizing bacteria (*Thioclava* and *Sulfurovum*) were identified for this system. Furthermore, the findings in investigation of thiosulfate formation and in the possible mechanisms were: 1) oxygen promoted Thio-S/TdS-S ratio from 4% to 24–26% with sulfate concentration kept above 15.2 mg S/L; 2) microbial activities were demonstrated to be crucial for ERATO; 3) oxygen contributed to thiosulfate formation through direct sulfide oxidation and, more importantly, indirect sulfide oxidation to S$^0$, which then reacts with accumulated sulfite which was caused by the lack of DsrC in BSR, upregulated by oxygen.

**Declaration of Competing Interest**

The authors declare that they have no known competing financial interests or personal relationships that could have appeared to influence the work reported in this paper.

**Acknowledgements**

This work was supported by the Hong Kong Innovation and Technology Commission (ITC-CNERC14EG03), the Research Grants Council of the Hong Kong SAR (T21-604/19-R), the International Science and Technology Cooperation Project of Huangpu District, Guang-

zhou (Grant No. 2020GH04), Ghent University (BOF/STA/202109/022), Korea Ministry of Environment (2022003050003).

**Tables**

**Table 1** Operation conditions for the bioreactor during days 0-349.

| Operating period (d) | Inf. DO (mg/L) | Inf. $NH_4^+$ (mg N/L) | HRT (min) | pH | T (°C) | DO (mg/L) |
|---|---|---|---|---|---|---|
| Phase I 0-136 | 2.7±0.5 | 28.3±1.3 | 40 | 7.2±0.1 | 27.8±0.5 | 0.3±0.2 |
| Phase II 137-270 | 0.5±0.3 | 29.5±1.5 | 40 | 7.2±0.1 | 27.1±0.7 | 0.4±0.2 |
| Phase III 271-349 | 3.6±0.8 | 28.9±1.4 | 40 | 7.1±0.1 | 27.0±0.5 | 0.4±0.2 |

Note: Inf. Refers to influent.

**Table 2** Conditions of batch test (BT) groups A and B.

| BT | $SO_4^{2-}$ (mg S/L) | $S^0$ (g/L) | TdS (mg S/L) | COD (mg/L) | $NH_4^+$ (mgN/L) | Air addition (mL) | Carriers | Initial pH |
|---|---|---|---|---|---|---|---|---|
| A1 | 50 | - | - | 110 | 30 | - | Alive* | 7.5 |
| A2 | 50 | - | - | 110 | 30 | 22.5 | Alive | 7.5 |
| A3 | 30 | - | - | 110 | 30 | 22.5 | Alive | 7.5 |
| A4 | 80 | - | - | 110 | 30 | 22.5 | Alive | 7.5 |
| A5 | 50 | - | - | 110 | 30 | 22.5 | Sterilized | 7.5 |
| B1 | - | - | 25 | 50 | 30 | 22.5 | Alive | 7.5 |
| B2 | - | - | 25 | 50 | 30 | 22.5 | Sterilized | 7.5 |
| B3 | 50 | 2.5 | - | 110 | 30 | - | Alive | 7.5 |

**Table 3** Performance of the bioreactor during days 0-349.

| Parameters | Phase I | Phase II | Phase III |
| --- | --- | --- | --- |
| Influent sulfate (mg S/L) | 54.0±3.7 | 48.1±5.9 | 50.6±2.7 |
| Effluent sulfate (mg S/L) | 33.4±5.7 | 24.9±3.9 | 25.0±3.2 |
| Sulfate decrease (mg S/L) | 20.6±4.5 | 23.2±6.3 | 25.6±2.2 |
| Effluent TdS (mg S/L) | 14.2±4.4 | 19.8±4.0 | 19.5±1.8 |
| Effluent thiosulfate (mg S/L) | 10.3±3.7 | 4.9±3.1 | 7.5±2.8 |
| Influent COD (mg S/L) | 118.4±11.6 | 109.8±6.7 | 110.4±5.0 |
| Effluent COD (mg S/L) | 49.6±12.0 | 59.3±8.6 | 57.3±5.0 |
| Sulfur balance ratio (%) | 119 | 106 | 105 |
| Volumetric BSR rate (kg S/(m$^3$·d)) | 0.74±0.16 | 0.84±0.23 | 0.92±0.09 |
| Volumetric TdS production rate (kg S/(m$^3$·d)) | 0.54±0.16 | 0.71±0.14 | 0.70±0.06 |
| Volumetric thiosulfate production rate (kg S/(m$^3$·d)) | 0.39±0.13 | 0.18±0.11 | 0.27±0.10 |
| Volumetric COD removal rate (kg/(m$^3$·d)) | 2.48±0.69 | 1.81±0.45 | 1.91±0.32 |
| Surface-specific BSR rate (g S/(m$^2$·d)) | 0.34±0.07 | 0.38±0.10 | 0.42±0.04 |
| Surface-specific TdS production rate (g S/(m$^2$·d)) | 0.25±0.07 | 0.32±0.07 | 0.32±0.03 |
| Surface-specific thiosulfate production rate (g S/(m$^2$·d)) | 0.18±0.06 | 0.08±0.05 | 0.12±0.05 |
| Surface-specific COD removal rate (g S/(m$^2$·d)) | 1.06±0.31 | 0.81±0.20 | 0.87±0.15 |
| Biofilm attached total solids (g/L) | 0.96±0.33 | 1.39±0.24 | 1.15±0.15 |
| Biofilm attached volatile solids (g/L) | 0.82±0.26 | 1.20±0.20 | 1.03±0.10 |
| Total suspended solids in effluent (mg/L) | 20.1±1.3 | 15.4±1.2 | 18.7±1.3 |
| Volatile suspended solids in effluent (mg/L) | 17.3±1.2 | 13.5±0.9 | 15.3±4.4 |

Note: the data were obtained in the end of each phase, 100-136d for Phase I, 247-270d for Phase II, and 301-349d for Phase III.

**Figures**

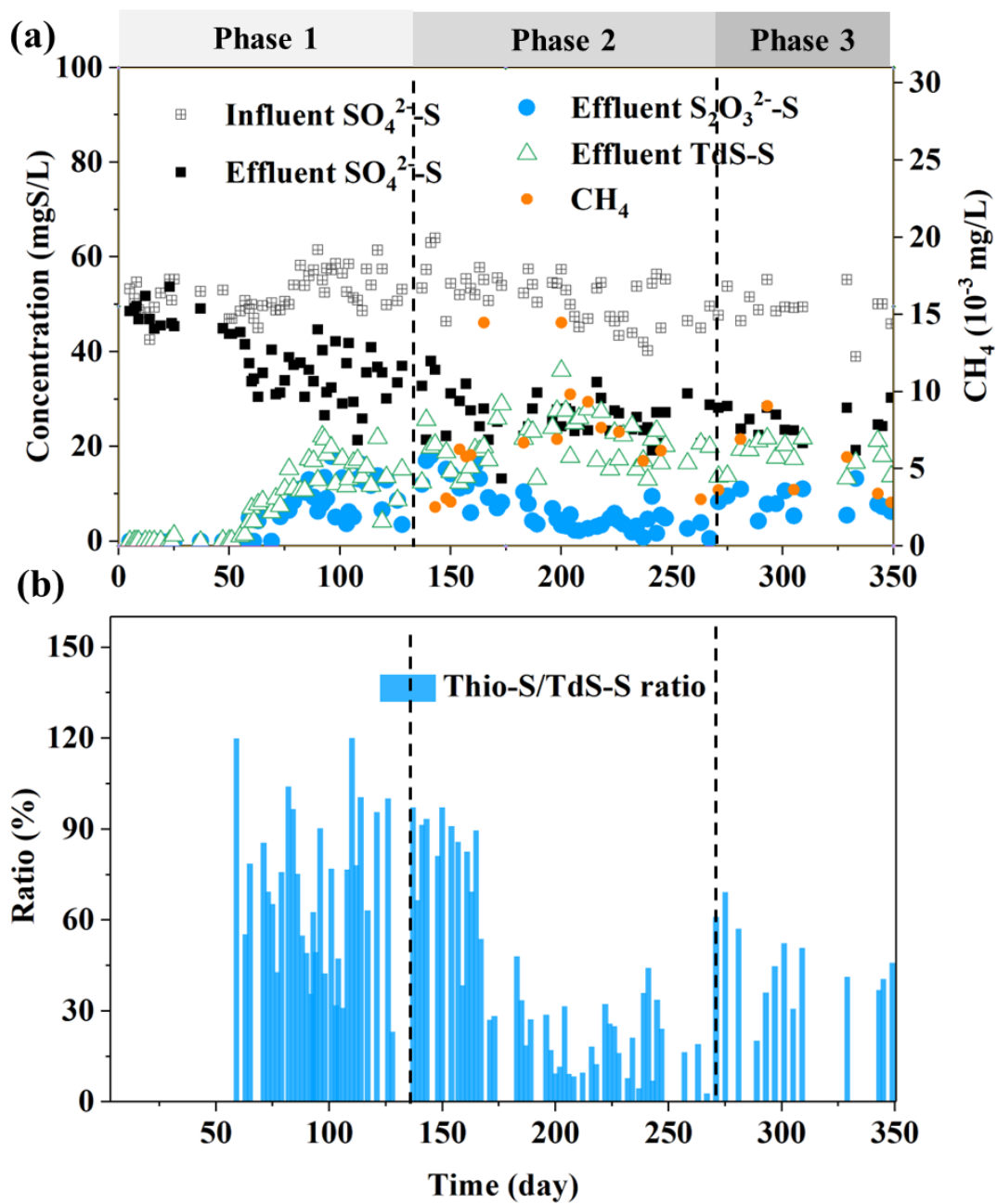

**Fig. 1** Performance of the bioreactor: (a) profile of sulfate, TdS, thiosulfate and methane; (b) changes in Thio-S/TdS-S ratio (note: TdS and Thio-S/TdS-S refer to total dissolved sulfide and thiosulfate-S$_{produced}$/TdS-S$_{produced}$, respectively).

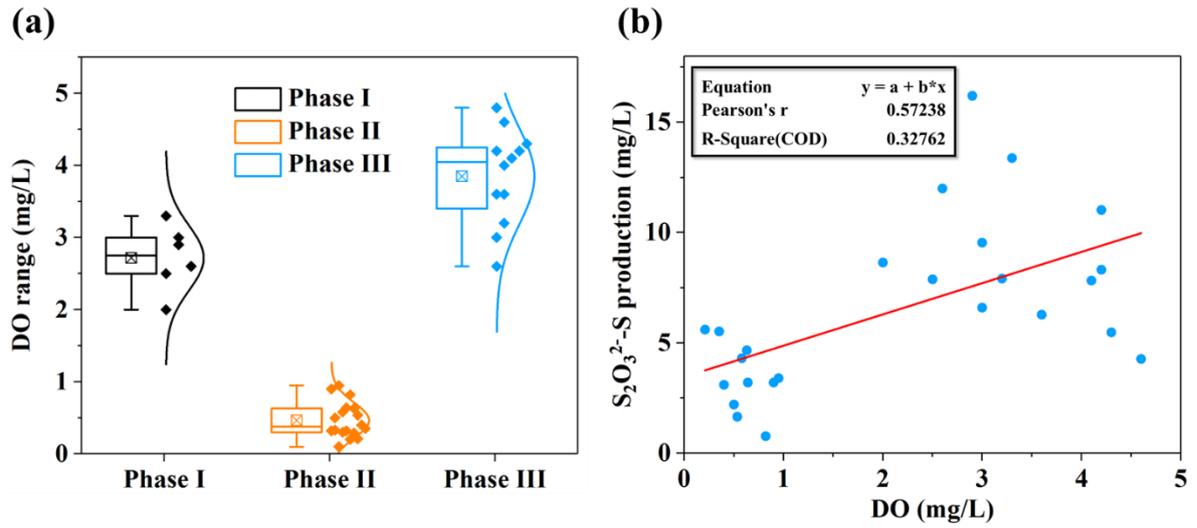

**Fig. 2** Profile of (a) boxplot for concentration of DO in influent and (b) liner relationship between the concentration of influent DO and thiosulfate.

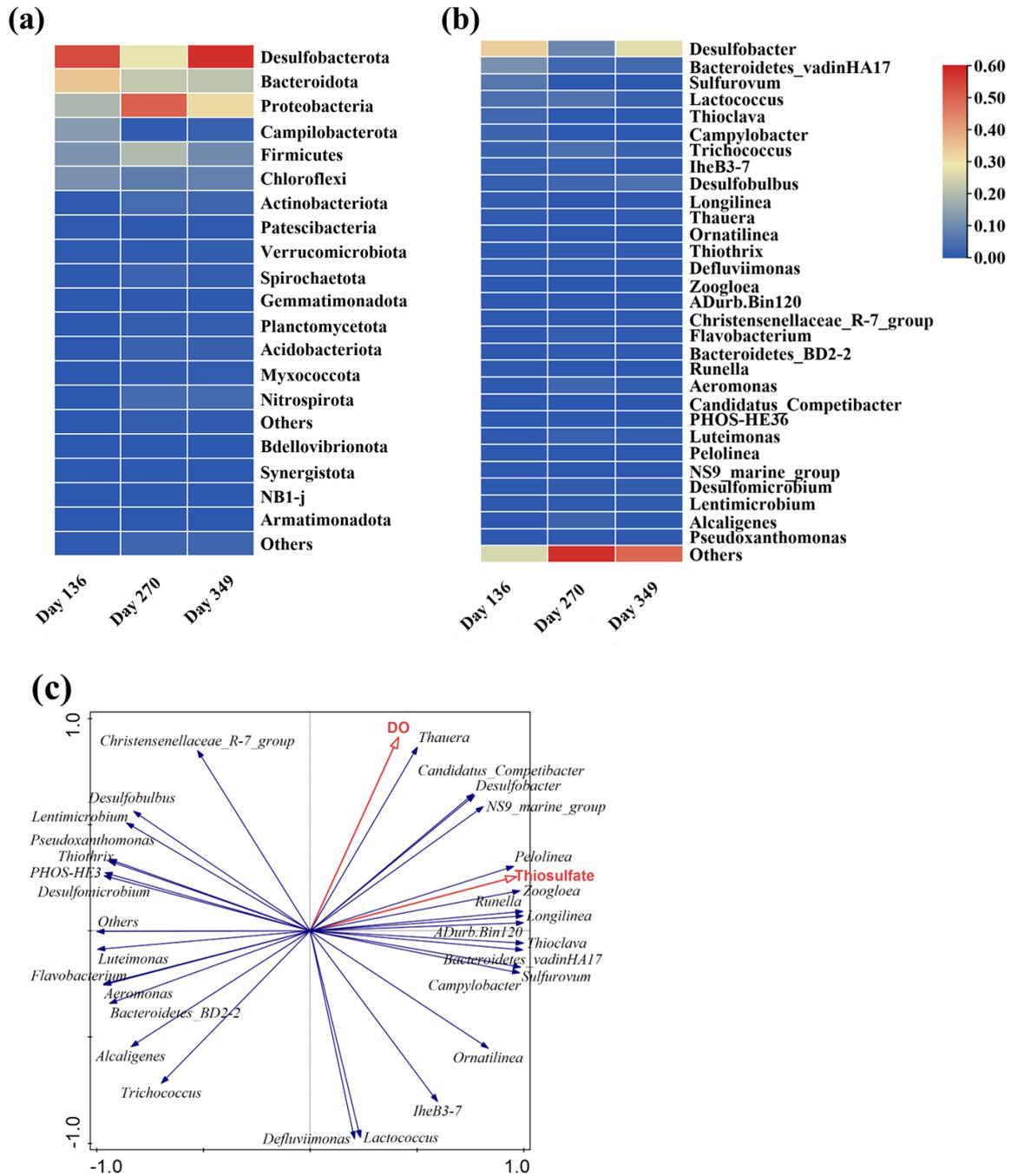

**Fig. 3** Dominant functional bacteria in the bioreactor during different periods at: (a) phylum and (b) genus level; (c) Redundancy analysis of the relationships between the concentration of influent DO as well as produced thiosulfate and dominant bacteria abundance at the genus level.

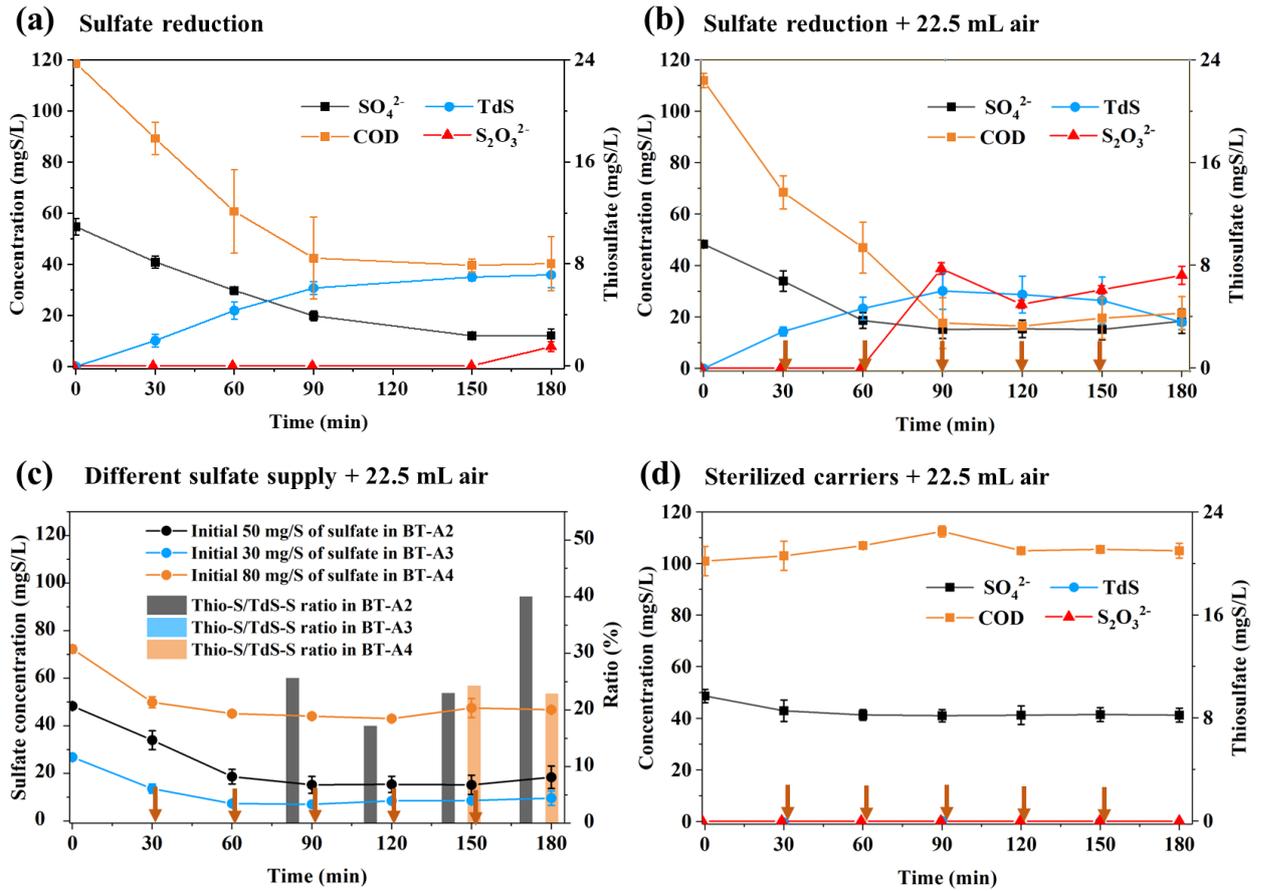

**Fig. 4** Results of BT-A: (a) BSR in BT-A1; (b) BSR with 22.5 mL air supply in BT-A2; (c) different initial sulfate concentrations in BT-A3 and A4 respectively (BT-A2 was also included for comparison); (d) 22.5 mL air supply with sterilized carriers in BT-A5 (note: a yellow arrow refers to 4.5 mL dosage of air, same in all BTs).

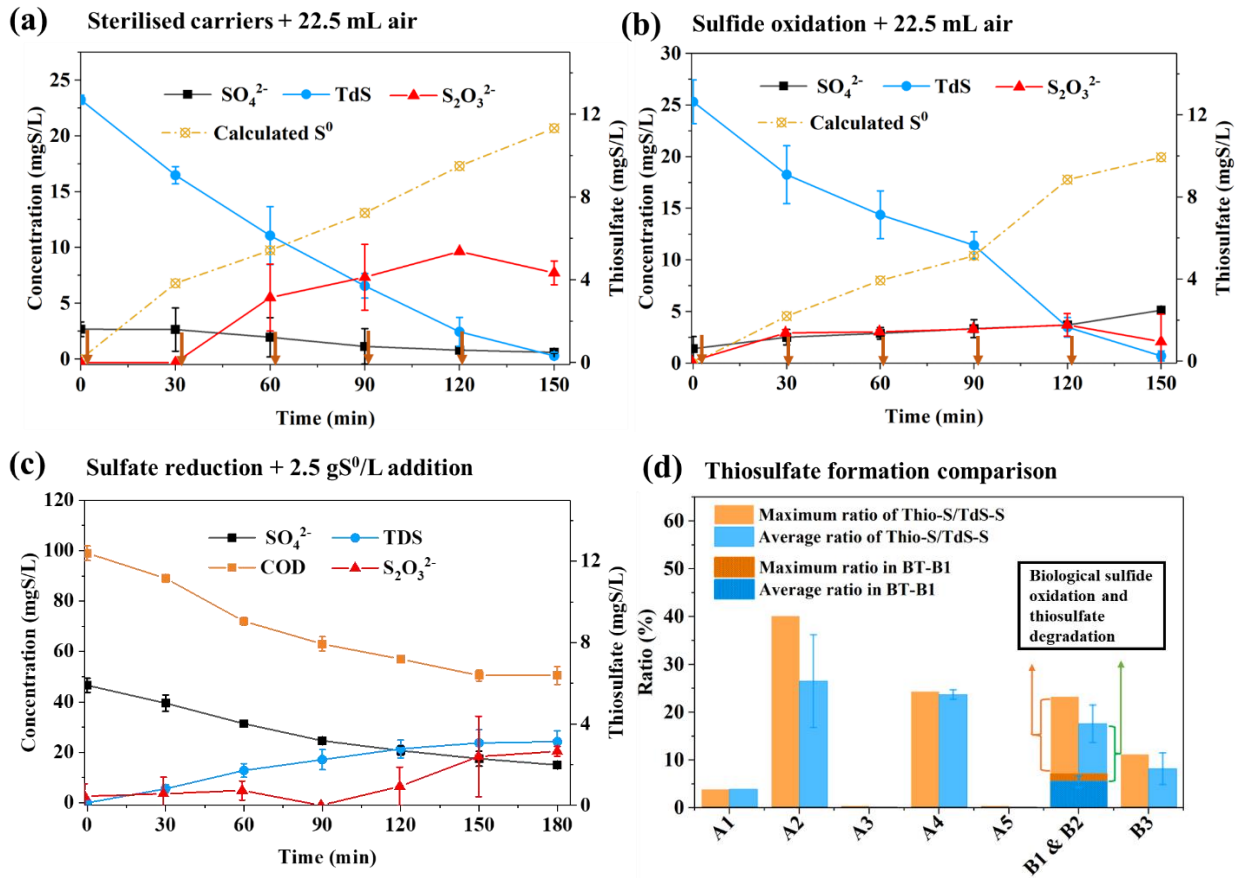

**Fig. 5** (a) Chemical sulfide oxidation with sterilized carriers in BT-B1; (b) sulfide oxidation with alive carriers in BT-B2; (c) BSR with external $S^0$ addition in BT-B3; (d) comparison of average and maximum ratios of Thio-S/TdS-S (note: the average ratio was calculated after thiosulfate produced in all BTs; the average ratio in BT-B1 and B2 was average ratio of thiosulfate production to initial TdS concentration).

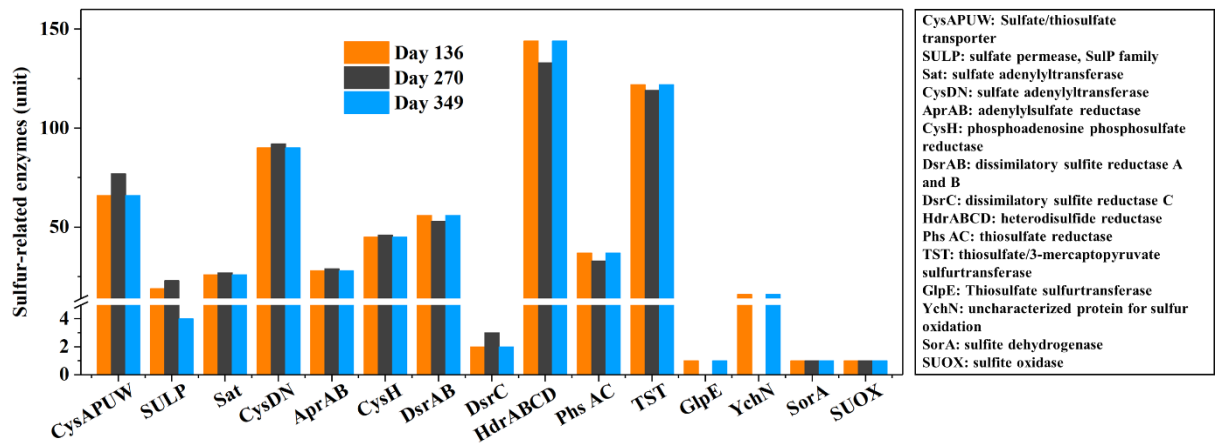

**Fig. 6** Predicted functional enzymes for sulfur conversion from PICRUSt.

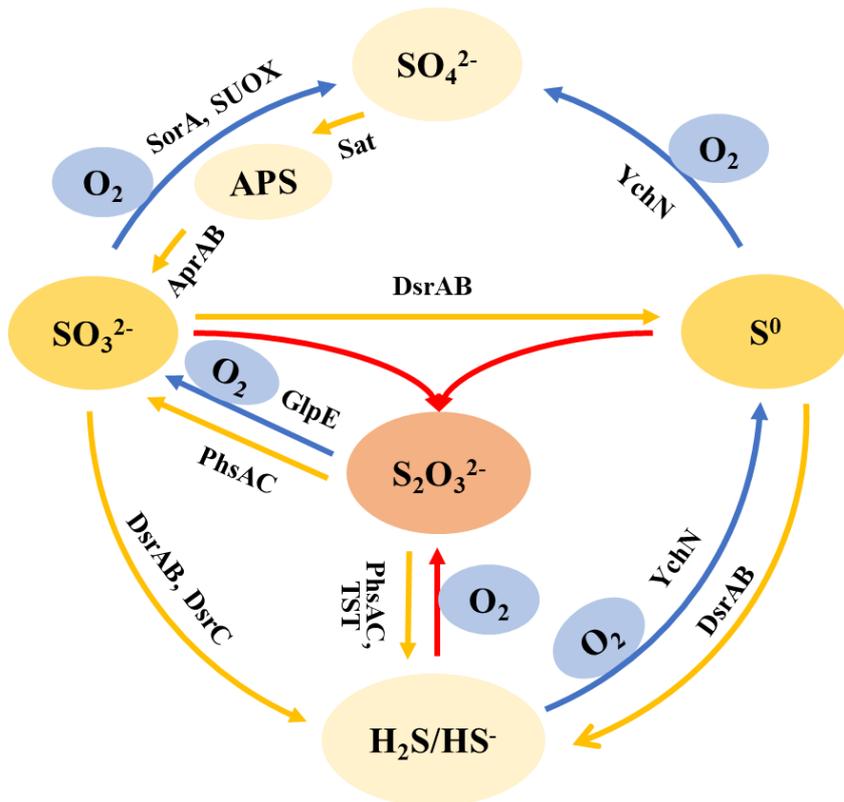

**Fig. 7** Proposed metabolism of thiosulfate conversion in ERATO (APS refers to adenosine phosphosulfate).

# SUPPLYMENTARY INFORMATION

# A new sulfur bioconversion process development for energy- and space-efficient secondary wastewater treatment

Chu-Kuan Jiang[a], Yang-Fan Deng[a,b], Hongxiao Guo[a], Guang-Hao Chen[a,b,*], Di Wu[a,c,d,*]

e. Department of Civil and Environmental Engineering, Water Technology Centre, Hong Kong Branch of Chinese National Engineering Research Centre for Control & Treatment of Heavy Metal Pollution, The Hong Kong University of Science and Technology, Hong Kong, China.

f. Wastewater Treatment Laboratory, Fok Ying Tung Graduate School, The Hong Kong University of Science and Technology, Guangzhou, China.

g. Centre for Environment and Energy Research, Ghent University Global Campus, Incheon, Republic of Korea.

h. Department of Green Chemistry and Technology, Ghent University, and Centre for Advanced Process Technology for Urban Resource Recovery, Ghent, Belgium.

*Corresponding authors:

Di Wu, Centre for Environment and Energy Research, Ghent University Global Campus (di.wu@ghent.ac.kr); Guang-Hao Chen, Department of Civil and Environmental Engineering, The Hong Kong University of Science and Technology (ceghchen@ust.hk).

**List of contents**

**SI 1:** Reactor setup and synthetic wastewater

**SI 2:** Stock solutions for batch experiments

**SI 3:** Kinetic analysis

**SI 4:** Raman analysis

**SI 5**: Extra BT for investigation of the effect of oxygen on BSR

**SI 6:** Microbial community analysis

**List of tables**

**Table S1** Diversity analysis of microbial samples.

**List of figures**

**Fig. S1** Setup of the lab-scale MBBR.

**Fig. S2** Raman spectroscopy analysis: images of (a) the biofilm sample and (b) the yellow complex; Raman spectra of (c) the biofilm sample and (b) the yellow complex.

**Fig. S3** Biological sulfate reduction with 35 mL air addition.

**SI 1: Reactor setup and synthetic wastewater**

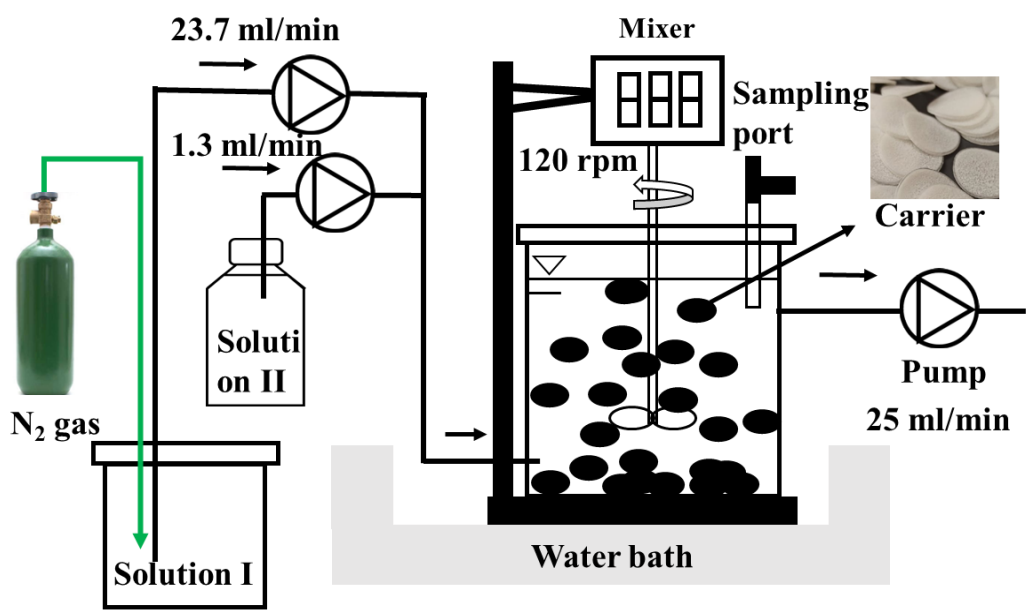

**Fig. S1** Setup of lab-scale MBBR (nitrogen gas flushing was applied in Phase II).

The composition of solution I was modified according to Wang et al. (2009). First, 22 mg TOC/L of glucose and 22 mg TOC/L of sodium acetate trihydrate (carbon ratio 1:1) were used as an organic source. The solution also contained 30 mg $NH_4^+$-N/L of ammonium chloride as an ammonia source, 270 mg/L sodium bicarbonate as an alkaline source, 10 mg/L yeast, 11.3 mg/L $KH_2PO_4$, 67 mg/L $MgCl_2·2H_2O$, 75 mg/L $CaCl_2·2H_2O$, and a 1 mL/L trace element solution (0.5 g/L EDTA, 0.1 g/L $FeSO_4$, 8 mg/L $ZnSO_4·7H_2O$, 4 mg/L $CoCl·6H_2O$, 20 mg/L $MnCl_2·4H_2O$, 5 mg/L $CuSO_4·5H_2O$, 4 mg/L $NaMoO_4·2H_2O$, 4 mg/L $NiCl_2·6H_2O$, 4 mg/L $NaSeO_4·10H_2O$, and 0.2 mg/L $H_3BO_4$). Solution II was 100% seawater obtained from the seawater toilet flushing water supply system. The flow rates of solutions I and II were 23.7 and 1.3 mL/min, respectively.

**SI 2: Stock solutions for batch experiments**

Nutrients were provided by freshly prepared solution I. Sulfur compounds and salinity (~1 gCl/L) were supplied by prepared stock solutions, including solutions of 1gS/L $Na_2S$, 1gS/L Na$_2$SO$_4$, 20gCl/L NaCl, and a mixed solution of 1 gS/L Na$_2$SO$_4$ and 20 gCl/L NaCl. Ultrapure water used for the stock solution was deoxygenated via nitrogen gas flushing for more than 60 min. Furthermore, 5–10 mL stock solutions were injected into sealed bottles with a syringe to achieve desired S and N compound concentrations.

**SI 3: Kinetic analysis**

Zero-order kinetics were observed for sulfate and organics degradation, and the surface-specific rates of BSR, organics degradation, and TdS production were described using a zero-order model. Thus, the rates were determined through the linear regression analysis of the concentrations of sulfate, TdS, and organics. The rates are shown below:

$$r_{SO4} = \frac{-dSO_4^{2-}}{Adt}$$

$$r_{\text{org}-C} = \frac{-dCOD}{Adt}$$

$$r_{\text{TdS}} = \frac{-dTdS}{Adt}$$

where $r_{SO4}$ (mg S/(m$^2$·h)), $r_{org-C}$ (mg COD/(m$^2$·h)) and $r_{TdS}$ (mg S/(m$^2$·h)) denote the surface-specific sulfate, organics utilisation rates and TdS production rate in BSR; SO$_4^{2-}$, COD, and TdS refer to the concentrations of each with the unit of mgS/L or mg COD/L; $A$ is the specific surface area of the biofilm in the MBBR (2.2 m$^2$/L) or the BT (1.0 m$^2$/L); $t$ is time (h).

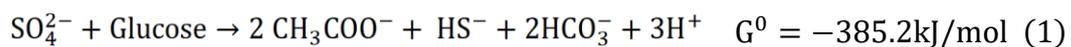

$$SO_4^{2-} + \text{Glucose} \rightarrow 2\ CH_3COO^- + HS^- + 2HCO_3^- + 3H^+ \quad G^0 = -385.2\text{kJ/mol} \quad (1)$$

$$SO_4^{2-} + CH_3COO^- \rightarrow HS^- + 2HCO_3^- \qquad \Delta G^0 = -47.6 \text{ kJ/mol} \quad (2)$$

**SI 4: Raman analysis**

The Raman spectra featured peaks at around 153, 214, and 474 cm$^{-1}$, corresponding to S$^0$ (McGuire et al., 2001). Thus, S$^0$ occurred in both the biofilm sample and the yellow complex, demonstrating that S$^0$ was produced in the MBBR. Compared with the biofilm samples, the peaks (assigned to S$^0$) of the yellow complex showed higher intensity and diversity, indicating that the yellow complex contained a much more S$^0$.

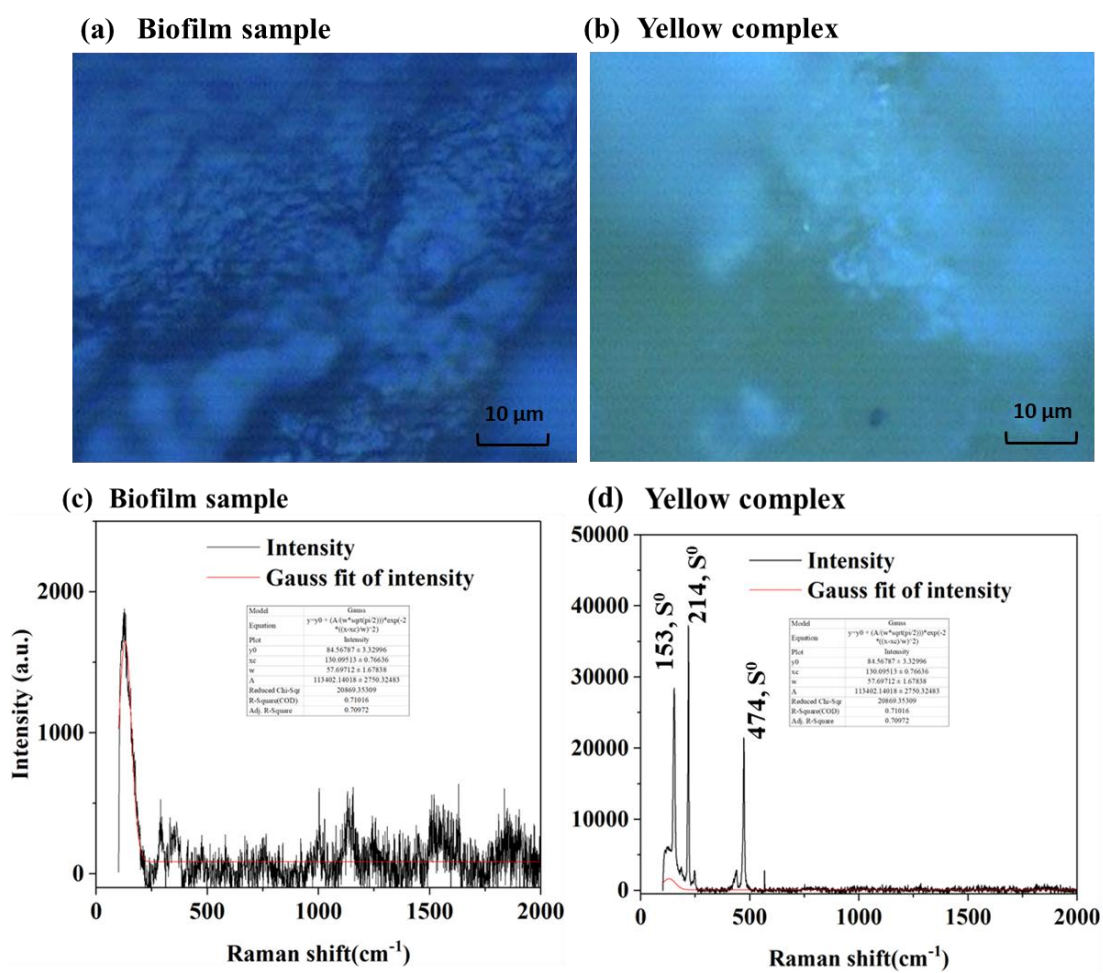

**Fig. S2** Raman spectroscopy analysis: images of (a) the biofilm sample and (b) the yellow complex; Raman spectra of (c) the biofilm sample and (b) the yellow complex.

## SI 5: Extra BT for investigation of the effect of oxygen on BSR

Compared to 22.5 mL air addition in BT-B, 35 mL of air was added to this batch. 7 mL air was dosed at 90, 120, 150, 180 and 210 min respectively. The initial conditions of this batch were set the same as BT-B. The results are presented in Fig. S3, 6.3 mgS/L thiosulfate was yielded at 120 min after the first dosage. During 120−180 min with extra 14 mL of air dosed, the sulfate and sulfide concentration fluctuated at 14.9−22.8 mgS/L and 23.0−30.2 mgS/L respectively. Meanwhile, thiosulfate-$S_{produced}$/sulfide-$S_{produced}$ decreased largely from 28% at 150 min to 17% at 180 min, indicating that suitable $O_{2-supply}/SO_4^{2-}{}_{initial}$ ratio (0.44) reached at 150 min. After 180 min when $O_{2-supply}/SO_4^{2-}{}_{initial}$ in range of 0.66−1.10, the increase and decrease of sulfate and sulfide respectively demonstrated the occurrence of sulfide oxidation and the excess supply of oxygen. The average of 5.5 ± 0.9 mgS/L thiosulfate produced in this batch was significantly less than the yield of BT-B (6.9 ± 0.4 mgS/L, $p < 0.05$).

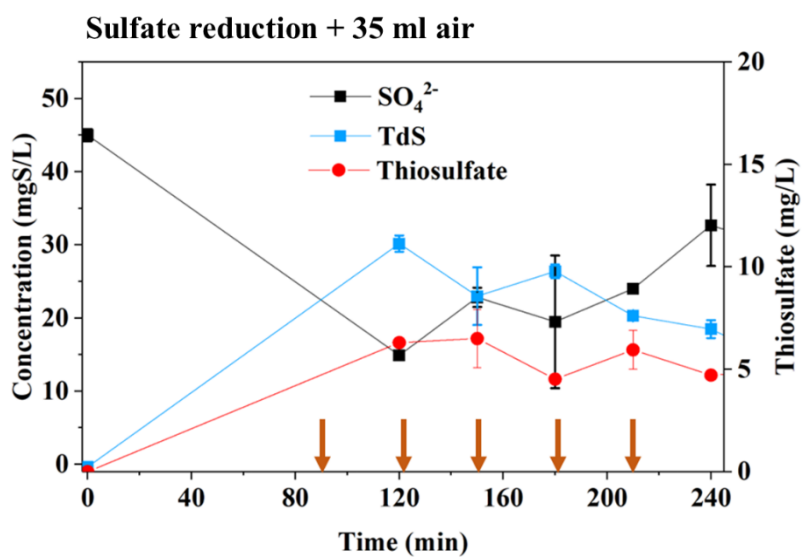

**Fig. S3** Biological sulfate reduction with 35 mL air addition (note: a yellow arrow refers to 7.0 mL dosage of air).

## SI 6: Microbial community analysis

### 6.1 Analytical method

Sludge samples were taken from carriers cultivated in the MBBR. Biomass samples were conserved with 50% ethanol solution for analysis via a 16S rDNA test at −36 °C (Li et al., 2018). Bacterial genomic DNA was extracted using the TIANamp Soil DNA Kit (Tiangen Biotech, Beijing) in accordance with the manufacturer's protocol. At least 60 ng DNA was extracted from each sample for further 16S rDNA gene PCR amplification and analysis via the next-generation sequencing technology (Novogene, Hong Kong). For result analysis, operational taxonomic unit (OTU) clustering was conducted. Taxonomic annotation was made for the representative sequence of each OTU to obtain the corresponding taxon information and taxon-based abundance distribution. Moreover, the OTUs were analyzed for alpha and beta diversity to obtain richness and evenness information in samples and common and unique OTU information among different samples.

### 6.2 Diversity analysis

As presented in Table S1, OTU change indicates microbial diversity in the MBBR. Over time, the OTUs first increased (from Phase I to II) and then decreased (from Phase II to III). Consistent with the OTU results, the Shannon, Simpson, Chao1, and Abundance-based coverage estimator (ACE) indices first increased and then decreased over time.

**Table S1** Diversity analysis of microbial samples.

| Sample | OTUs | Shannon | Simpson | Chao1 | ACE | Coverage |
|---|---|---|---|---|---|---|
| Day 136 | 863 | 5.223 | 0.914 | 949.141 | 956.159 | 0.998 |
| Day 270 | 2693 | 7.958 | 0.983 | 3271.422 | 3381.311 | 0.988 |
| Day 349 | 2507 | 7.288 | 0.933 | 3032.626 | 3112.940 | 0.989 |

Note: operational taxonomic units (OTUs) were classified with a sequence similarity of over 0.97.

**SI References**